\documentclass[onecollarge,runningheads]{svjour2}
\smartqed  
\usepackage{graphicx}
\journalname{Journal of Astrophysics and Astronomy}
\begin{document}
\title{
Anisotropic Nature of  Work Function in a Strong Quantizing Magnetic
Field}
\titlerunning{Anisotropic Nature of Work Function in Strong
Quantizing Magnetic Field}
\author{Arpita Ghosh and Somenath Chakrabarty}
\institute{
Department of Physics, Visva-Bharati\\ Santiniketan 731 235,\\ 
West Bengal, India\\ E-Mail:somenath.chakrabarty@visva-bharati.ac.in}
\date{Received: date / Accepted: date}
\maketitle
\begin{abstract}
Following an extremely interesting idea \cite{R1}, published long ago, the work
function associated with the emission of ultra-relativistic electrons from magnetically 
deformed metallic crystal of astrophysical relevance is obtained using 
relativistic version of Thomas-Fermi type model. In the present scenario, surprisingly, 
the work function becomes anisotropic; the longitudinal part is an increasing function of 
magnetic field strength, whereas the transverse part diverges.
\end{abstract}
\section{Introduction}
In condensed matter physics the work function is the 
minimum energy needed to remove an electron from a solid to
a point immediately outside the solid surface or equivalently, the energy 
needed to move
an electron from the Fermi level into vacuum. Here the word immediately
means
that the final electron position is far from the surface on the atomic
scale but still close to the solid on the macroscopic scale. It is also
known that the work
function is a characteristic property for any solid face of a
substance with a conduction band, which may be empty or partly filled. For a
metal, the Fermi level is inside the conduction band, indicating that
the band is partly filled. For an insulator, however, the Fermi level 
lies within
the band gap, indicating an empty conduction band; in this case, the
minimum energy to remove an electron is about the sum of half the band
gap, and the work function.

In the free electron model the valence electrons move freely inside the 
metal but find a confining potential step, say $C$ at the
boundary of the metal. In the system's ground state, energy levels below
the Fermi energy are occupied, and those above the Fermi Level
are empty. The energy required to liberate an electron in the
Fermi Level is the work function and is given by $W_f=C-\mu_e$, where
$\mu_e$ is the Fermi energy. Therefore,
the work function of a metal is usually defined as the smallest energy needed to
extract an electron at zero temperature. Formally, this definition is made for 
an infinitely large crystal plane in which one takes an electron from infinitely
deep inside the crystal and brings it through the surface, infinitely
far away into the vacuum. The exact definition of work function, which is valid
in the atomic scale is the work done in bringing an electron from far below 
the surface,
compared to atomic dimensions but not far compared to crystal
dimensions.

The work function of metals varies from one crystal plane to another and
also varies slightly with temperature. For a metal, the work function
has a simple interpretation. At absolute zero, the energy of the most
energetic electrons in a metal is referred to as the Fermi energy; the
work function of a metal is then equal to the energy required to raise
an electron with the Fermi energy to the energy level corresponding to
an electron at rest in vacuum. The work function of a semiconductor or
an insulator, however, has the same interpretation, but in these materials the
Fermi level is in general not occupied by electrons and thus has a more
abstract meaning (see \cite{sn,fr,tb,sk,ej1,jb} as some of the
fundamental papers on work function).

The work function is associated with three types of electron
emission processes: photo-emission, thermionic emission and field
emission. All of them have a large number of applications in various branches of science
and technology; starting from electronic valves, a very old type
electronic device, to modern opto-electronic devices used to convert optical 
signals to
electrical signals etc. The other important applications are in  
photo-multiplier
tube (PMT), CCD, etc. In the recent years, it is found that the field emission 
process has a lot of important applications in modern nano technology.

Similar to the down to earth application of work function in condensed
matter physics and its various scientific and technological
applications, the work function also plays vital role in many
astrophysical processes, e.g., in the formation of magneto-sphere by the
emission of electrons from the polar region induced by strong
electrostatic field in
strongly magnetized neutron stars. The present investigation is mainly
associated with the emission of high energy electrons from the dense
metallic iron crystal present at the crustal region of strongly
magnetized neutron stars.  
The study of the formation of plasma in a pulsar magnetosphere is a quite old 
but still an unresolved astrophysical issue. In the formation of 
magneto-spheric 
plasma, it is generally assumed that there must be an initial high energy 
electron flux from the magnetized neutron stars. At the poles of a neutron 
star the emitted charged particles flow only along the magnetic field lines. 
Further a rotating magnetized neutron star generates extremely high 
electro-static potential difference at the poles.
The flow of high energy electrons along the direction of magnetic field
lines and their penetration through the light cylinder is pictured with the 
current carrying conductors. Naturally, if the conductor is broken near the 
pulsar surface the entire potential difference will be developed across a thin 
gap, called polar gap. This is of course based on the assumption that
above a critical height from the polar gap, because of high electrical 
conductivity of the plasma, the electric field $E_{\vert\vert}$, parallel with the 
magnetic field near the poles is quenched. Further, a steady acceleration of electrons 
originating at the polar region of neutron stars, travelling along the field lines, 
will produce magnetically convertible curvature $\gamma$-rays.
If these curvature $\gamma$-ray photons have energies $>2m_ec^2$ (with $m_e$ 
the electron rest mass and $c$ the velocity of light), then pairs of $e^--e^+$ 
will be produced in enormous amount with very high efficiency near the polar 
gap. These produced $e^--e^+$ pairs form what is known as the magneto-spheric 
plasma \cite{R2,R3,R4,R5,R6,R7,R8}.  
The emission of electrons from the polar region of neutron stars is mainly dominated by the
cold emission or the field emission \cite{R12}, driven by electro-static force at the poles, produced by 
the strong magnetic field of rotating neutron stars. The work function
plays a major role in the
emission processes. Moreover, it is interesting to note that electrons
are only emitted from the polar region of magnetized neutron stars, but
never from some region far from the poles.
A large number of theoretical and numerical techniques were developed to obtain analytical expression and
also the numerical values of work functions of various 
materials for which the experimental values are also known. Self-consistent jellium-background model, 
embedding atom-jellium model, density functional theory (DFT) etc., were used to obtain numerical 
values for work functions \cite{ek,wc,mvvv}. In a recent article the work function has been 
identified with the exchange energy of electrons within the material \cite{hph,pl,rs}.

The process of extracting electrons from the outer crust region of strongly 
magnetized neutron stars, including the most exotic stellar objects, the 
magnetars, it requires a more or less exact description of the structure of matter in that region. 
From the knowledge of structural deformation of atoms in strong magnetic field; we expect that 
the departure from spherical nature to a cigar shape allows us to assume that
because of high density the electron distribution around the iron nuclei at the
outer crust region may be 
replaced by Wigner-Seitz type cells of approximately cylindrical in structure \cite{R9,R10,R11}. 
We further assume that the electron gas inside the cells are strongly degenerate and at zero 
temperature (In presence of strong quantizing magnetic field the
atoms get deformed from their usual spherical shape and also get
contracted, then it is quite natural that the electron density becomes so
high inside the cells that their chemical potential $\mu_e \gg T$, the temperature of the
degenerate electron gas. Then the temperature of the system may be assumed to
be very close to zero). It is well known that the presence of extraordinarily large magnetic field not only
distorts the crystalline structure of dense metallic iron, also significantly modifies the 
electrical properties of such matter. As for example, the electrical conductivity, 
which is otherwise isotropic,
becomes highly anisotropic in presence of strong quantizing magnetic field \cite{pbh,po,py}. 
In presence of strong magnetic field, iron crystal is highly conducting in the direction 
parallel to the magnetic field, whereas flow of current in the perpendicular direction is 
severely inhibited. 

The aim of this article (i) is to show that the work function 
associated with the emission of electrons along the direction of
magnetic field at the polar region of strongly magnetized neutron 
stars increases with the strength of magnetic field and (ii) it will be
shown 
that the same quantity associated with the emission of electrons in the
direction transverse to the magnetic field direction is infinitely large.
Therefore the main result of this article is to show that in presence of
a strong quantizing magnetic field, the work function 
becomes anisotropic (see also \cite{pl,rs} and \cite{md1,md2}).
We have noticed that the result does not depend on the dimension of the crustal matter. 
The scenario is very much analogous to the charge transport 
mechanism within the metal in presence of strong magnetic field. The electron emission process may be
assumed to be a kind of charge transport process from inside the metal
to outside. We shall show in this article
that analogous to the
internal charge transport, this is also anisotropic in presence of strong 
magnetic field and if the magnetic field strength is high enough,
the emission process in the transverse direction is totally forbidden.
To the best of our knowledge, using this simple idea for cold cathode emission 
\cite{R1}, the study of anisotropic nature of work function in presence of strong quantizing magnetic field, 
relevant for strongly magnetized neutron star crustal region, has not been studied earlier. 
Of course, a completely different approach, popularly known as density functional theory (DFT) 
has been used by Lai et. al. to study properties of dense neutron star crustal matter. 
They have also obtained the work function as a function of magnetic
field \cite{md1,md2,ad} (see also \cite{lk}). 
Surprisingly, our simple minded model gives almost same kind of magnetic field dependence for 
the work function as obtained by Lai et. al. 

We have organized this article in the following manner: in section 2, we shall
obtain an approximate analytical expression for work function associated with the emission of 
electrons along the field direction. In section 3, we shall obtain the same quantity 
corresponding to the emission transverse to the direction of magnetic field, and show that
this particular component is infinitely large. 
Whereas in the last section we
shall give conclusions and future prospects of this work.
\section{Longitudinal Emissions}
In this section, for the sake of completeness, we shall repeat some of the derivations presented in
\cite{R1}. Based on this particular work we shall develop a formalism to obtain the work function 
associated with the emission of electrons along the direction of magnetic field from dense crystalline
structure of mainly metallic iron in the outer crust region of neutron stars. 
It should be noted that the physical picture of our model is completely 
different from \cite{R1}.
In our model calculation the arrangement of cylindrically deformed WS cells in the crustal region  
are such that the field lines 
and the axes of the cylinders are parallel with each other. 
To obtain the work function for both axial and transverse emission processes, 
let us consider fig.(1), where we have considered a cylindrically
deformed WS cell of the crustal matter.
Here the magnetic field is along z-axis and r-axis is orthogonal to the direction of 
magnetic field lines, and we assume azimuthal symmetric geometrical structure.  
To get the longitudinal part of the work function, we assume that an electron has come out 
from the cylindrically deformed cell through one of the plane faces, and is at the point $P$. 
The electro-static potential at $Q$ produced by this electron is given by  
\begin{equation}
\phi^{(p)}(r,z)=\frac{e}{[r^2+(z-h)^2]^{1/2}}
\end{equation}	
where $AP=h$, $PD=r$ and $CQ=z$. This equation can also be expressed in the following form 
\begin{eqnarray}
\phi^{(p)}(r,z)=e\int_{\xi=0}^\infty J_0(\xi r)&&\exp[\mp(z-h)\xi]d\xi\nonumber \\ && ~~~~{\rm{for}}~~ z> ~~{\rm{or}}~~
<h
\end{eqnarray}
For the charge distribution within the distorted WS cell, the Poisson's equation is given by 
\begin{equation}
\nabla^2\phi=4\pi en_e
\end{equation}
where $\phi$ is the electro-static potential and $n_e$ is the electron density, given by 
\begin{equation}
n_e=\frac{eB}{2\pi^2}\sum_{\nu=0}^{\nu_{max}}(2-\delta_{\nu 0})(\mu^2-m^2-2\nu eB)^{1/2}
\end{equation}
where $\nu$ is the Landau quantum number with $\nu_{max}$ the upper limit (it is finite for $T=0$).
The factor $2-\delta_{\nu 0}$ within the sum takes care of singly degenerate zeroth Landau level 
and doubly degenerate other levels with $\nu\neq 0$
and $\mu$ is the chemical potential for the electrons, given by 
$\mu=E_F-e\phi (r)={\rm{constant}}$,
this is the so called Thomas-Fermi condition, $E_F$ is the Fermi energy for the electrons, given by
$E_F=(p_F^2+m_e^2)^{1/2}$ (throughout
this article we have chosen $\hbar=c=1$). 
Hence we can re-write the Poisson's equation in the following form
\begin{equation}
\nabla^2\phi=\frac{2e^2 B}{\pi}\sum_{\nu=0}^{\nu{max}}(2-\delta_{\nu 0})[(\mu+e\phi)^2-m_e ^2-2\nu
eB]^{1/2}
\end{equation}
To get an analytical solution for $\phi$ in cylindrical coordinate system $(r,\theta,z)$, with
azimuthal symmetry, we put, $m_e=0$, i.e., the kinetic energy is assumed to be high enough
compared to the electron rest mass and $\nu_{max}=0$. The later is a valid approximation, 
provided the magnetic field is too high ($\geq 10^{14}$G), so that the electrons occupy only 
their zeroth Landau level. We expect, that the above assumptions are not so drastic for the type of physical 
system considered here.
Under such approximation, the Poisson's equation reduces to 
\begin{equation}
\nabla^2 \phi \approx \frac{2e^2B}{\pi}(\mu+e\phi)
\end{equation}
Now defining $\psi=\mu+e\phi$ the Poisson's equation becomes 
\begin{equation}
\nabla^2\psi \approx \frac{2e^3B}{\pi}\psi=\lambda^2\psi
\end{equation}
In cylindrical coordinate $(r,\theta,z)$, the above equation can be written as 
\begin{equation}
\frac{\partial^2\psi}{\partial r^2}+\frac{1}{r}\frac{\partial\psi}{\partial
r}+\frac{\partial^2\psi}{\partial^2z}=\lambda^2\psi
\end{equation}
The solution in the cylindrical coordinate system is given by 
\begin{eqnarray}
\psi(r,z)=\int_{\xi=0}^\infty J_0(\xi r)a(\xi)&&\exp[(\xi^2+\lambda^2)^{1/2}z]d\xi\nonumber \\ &&
~~{\rm{with}}~~ z<0
\end{eqnarray}
Here $\lambda^2=2e^3 B/\pi$ (in this connection, one should not that the parameter $\lambda$ 
used in \cite{R1} has completely different physical meaning) and $a(\xi)$ is some unknown 
spectral function. Then following \cite{R1},
we assume that there exist a fictitious secondary field in vacuum. The work function is defined as the work 
done by this field in pulling out an electron from just inside the material surface 
to infinity. The secondary field, as introduced in \cite{R1}, is expressed in coherent 
with $\phi^{(p)}(r,z)$ and $\phi(r,z)$, and is given by
\begin{eqnarray}
\phi^{(s)}(r,z)=e\int_{\xi =0}^\infty J_0(\xi r)f(\xi)&&\exp(-\xi z)d\xi \nonumber \\ &&~~{\rm{with}}~~ z>0
\end{eqnarray}
where $f(\xi)$ is again some unknown spectral function. 
To obtain $f(\xi)$, we follow \cite{R1} and use the continuity conditions for tangential and transverse 
components of electric field and the displacement vector respectively, given by
\begin{equation}
E^{(t)}=E^{(p)t}+E^{(s)t}~~{\rm{and}}~~
D^\perp=D^{(p)\perp}+D^{(s)\perp}
\end{equation}
where for the case of electron emission along magnetic field direction, 
the tangential part of electric field components are given by $r$ derivatives of 
the corresponding potentials with a negative sign as the multiplicative factor, e.g., 
$E^{(t)}=-\partial \phi/\partial r$ and similarly for others. The axial part of the
components for displacement vectors are obtained by taking $z$ derivatives of the
corresponding potentials multiplied by $-1$ and the dielectric constant of the medium, which are
assumed to be unity, i.e., $D^\perp=-K\partial \phi/\partial z=-\partial \phi/\partial z=E^\perp$.  
Using the $z$ derivatives for exponential functions and the relation 
$\partial J_0(\xi r)/\partial r= \xi J_0^\prime(\xi r)= -\xi J_1(\xi r)$ for the $r$
derivatives for Bessel functions, we finally have following \cite{R1}
\begin{equation}
f(\xi)=\left [\frac{1-\left (1+\frac{\lambda^2}{\xi^2}\right )^{1/2}}
{1+\left (1+\frac{\lambda^2}{\xi^2}\right )^{1/2}}\right] \exp(-\xi h)
\end{equation}
Then we have the secondary potential 
\begin{equation}
\phi^{(s)}(r,z)=e\int_{\xi=0}^\infty J_0(\xi r)\left [\frac{1-\left (1+\frac{\lambda^2}{\xi^2}\right )^{1/2}}
{1+\left (1+\frac{\lambda^2}{\xi^2}\right )^{1/2}}\right] \exp [-\xi(h+z)] d\xi ~~{\rm{for}}~~z>0
\end{equation}
Hence the secondary field along axial direction at $r=0$ is given by 
\begin{eqnarray}
E_z^{(s)}&=& -\frac{\partial \phi^{(s)}}{\partial z} \\
&=&e\int_{\xi=0}^\infty \left [ \frac{1-\left (1+\frac{\lambda^2}{\xi^2}\right )^{1/2}}
{1+\left (1+\frac{\lambda^2}{\xi^2}\right )^{1/2}}\right] \exp [-\xi(h+z)] \xi d\xi 
\end{eqnarray}
Then the force acting on an electron at $z=h$, which is at the verge of emission, is given by 
\begin{equation}
F_z^{(s)}=e E_z^{(s)}=e^2\int_{\xi=0}^\infty \left [\frac{1-\left (1+\frac{\lambda^2}{\xi^2}\right )^{1/2}}
{1+\left (1+\frac{\lambda^2}{\xi^2}\right )^{1/2}}\right] \exp (-2h\xi) \xi d\xi
\end{equation}
The work done in pulling out an electron from just inside the metal surface to infinity is then given by
\begin{eqnarray}
W_f&=&-\int_0^\infty F_z^{(s)}dh \\
&=&-\frac{e^2}{2}\int_{\xi=0}^\infty \frac{1-\left (1+\frac{\lambda^2}{\xi^2}\right )^{1/2}}
{1+\left (1+\frac{\lambda^2}{\xi^2}\right )^{1/2}} d\xi
\end{eqnarray}
Substituting $\xi=\lambda \sinh u$, we have 
\begin{equation}
W_f=\frac{e^2}{2}\lambda\int_0^\infty \exp(-2u)\cosh u~ du=\frac{\lambda}{3}e^2=\left
(\frac{2e^3B}{\pi}\right )^{1/2}e^2
\end{equation}
Expressing the magnetic field in terms of the critical field strength, we get
\begin{equation}
W_f=\frac{\lambda}{3}e^2=\frac{1}{3}\left (\frac{2B}{\pi B_c^{(e)}}\right )^{1/2}m_ee^3
\end{equation}
where $B_c^{(e)}\approx 4.43\times 10^{13}$G, the typical field strength for electrons to populate
their Landau levels in the relativistic scenario. This equation gives the variation of work function with the strength of magnetic field ($\sim B^{1/2}$) associated with the emission of electrons along the direction of
magnetic field. From this expression it is also obvious that for a given magnetic field strength, 
if $m_e$ is replaced by $m_p$, the proton mass or $m_I$, mass of the ions, then, 
since $m_I\gg m_p\gg m_e$, 
the work functions associated with their emissions along $z$-direction
can also be expressed by
the same kind of inequalities. Then obviously, it needs high temperature for thermo-ionic emission of protons or ions, high
electric field for their field emissions and high frequency incident photons are essential for the
photo-emission of these heavier components. Since the work function
$\propto B^{1/2}$, the emission of electrons should decrease with the
increase in magnetic field strength. Physically, it means that the electrons become more
strongly bound within the metals. 
In fig.(2) we have shown the variation of $W_f$, the work function associated with the emission of
electrons along the magnetic field direction, with $B$, the strength of magnetic field. 
This graph clearly shows that the work function is an increasing function of magnetic field strength
and for low field values, $W_f$ is a few ev in magnitude, 
which is of the same orders of magnitudes with the experimentally known values for laboratory metals. 
Since in this approximate calculation, the magnetic field is
assumed to be extremely high, we are therefore unable to extrapolate 
this model to very low field values. Also,
in this model we can not show the variation of $W_f$ from one kind of metal to another. However, a
comparison of our result with fig.(4) of \cite{md1} shows that in the DFT calculation also the variation 
of work function from one kind of metal to another one is not so significant. Further, our result is
consistent with DFT calculation.
\section{Transverse Emissions}
Let us now consider the emission of electrons in the transverse direction. As shown in fig.(1),
here $p$ is the position of an electron, came out through the curved surface, then 
the electro-static potential at $q$ is given by 
\begin{equation}
\phi^{(p)}(r,z)=\frac{e}{[z^2+(r-r_0)^2]^{1/2}}
\end{equation}
where $ap=r_0$, $cq=r$ and $pd=z$. It can also be expressed as 
\begin{eqnarray}
\phi^{(p)}(r,z)=e\int_{\xi =0}^\infty &&J_0[\xi(r-r_0)]\exp{(\mp z\xi)}d\xi \nonumber \\ &&~~{\rm{for}}~~ z>0
~~{\rm{or}}~~ z<0
\end{eqnarray}
whereas the form of $\phi(r,z)$ (or the modified form $\psi(r,z)$) and $\phi^{(s)}(r,z)$ remain unchanged 
(eqns.(6)-(7)). 
Using the relation 
\begin{equation}
J_0(u-v)=\sum_{k=-\infty}^{+\infty}J_k(u)J_k(v)
\end{equation}
we can express
\begin{equation}
J_0(\xi r)=J_0\left [\xi(r+r_0-r_0)\right]=J_0\left [\xi(r+r_0)-\xi
r_0\right]=\sum_{k=-\infty}^{+\infty} J_k\left [\xi(r+r_0)\right ]J_k(\xi r_0)
\end{equation}
Then we have
\begin{equation}
\psi(r,z)=\sum_{k=-\infty}^{+\infty}J_k(\xi r_0)\int_{\xi=0}^\infty J_k\left [\xi(r+r_0)\right]
a(\xi)\exp[-(\xi^2+\lambda^2)^{1/2}\mid z\mid] d\xi
\end{equation}
Similarly we have
\begin{equation}
\phi^{(p)}(r,z)=e\sum_{k=-\infty}^{+\infty}J_k(2\xi r_0)\int_{\xi=0}^\infty J_k[\xi(r+r_0)]
\exp(-\xi\mid z\mid) d\xi
\end{equation}
and 
\begin{equation}
\phi^{(s)}(r,z)=e\sum_{k=-\infty}^{+\infty}J_k(\xi r_0)\int_{\xi=0}^\infty J_k[\xi(r+r_0)]f(\xi)
\exp(-\xi\mid z\mid) d\xi
\end{equation}
Then from the continuity conditions on the curved surface along $r$ direction, we have 
\begin{eqnarray}
&&\sum_{k=-\infty}^{+\infty}J_k(\xi r_0)\int_{\xi=0}^\infty J_k^\prime[\xi(r+r_0)]\xi
a(\xi)\exp[-(\xi^2+\lambda^2)^{1/2}\mid z\mid] d\xi \nonumber \\
&=&e\sum_{k=-\infty}^{+\infty}J_k(2\xi r_0)\int_{\xi=0}^\infty
J_k^\prime[\xi(r+r_0)]\xi\exp(-\xi\mid z\mid) d\xi \nonumber \\
&+&e\sum_{k=-\infty}^{+\infty}J_k(\xi r_0)\int_{\xi=0}^\infty J_k^\prime[\xi(r+r_0)]\xi
f(\xi)\exp(-\xi\mid z\mid) d\xi 
\end{eqnarray}
Putting $z=0$, $r=R$ and redefining the spectral function $a(\xi)=a(\xi)/e$, we have
\begin{eqnarray}
&&\int_{\xi=0}^\infty \xi d\xi\sum_{k=-\infty}^{+\infty}J_k^\prime[(R+r_0)\xi]\{a(\xi)
J_R(\xi r_0)\nonumber \\&-&J_k(2\xi r_0)-J_k(\xi r_0)f(\xi)\}=0
\end{eqnarray}
Since $\xi d\xi$ is arbitrary, we have
\begin{eqnarray}
&&\sum_{k=-\infty}^{+\infty}J_k^\prime[(R+r_0)\xi]\{a(\xi)
J_R(\xi r_0)\nonumber \\&-&J_k(2\xi r_0)-J_k(\xi r_0)f(\xi)\}=0
\end{eqnarray}
Similarly the continuity condition along $z$-direction on the curved surface at $z=0$ and $r=R$ is given by
\begin{eqnarray}
&&\int_{\xi=0}^\infty\sum_{k=-\infty}^{+\infty}J_k(\xi r_0)J_k[(R+r_0)\xi]a(\xi)(\xi^2+\lambda^2)^{1/2}d\xi \nonumber \\
&=&\int_{\xi=0}^\infty\sum_{k=-\infty}^{+\infty}J_k(2\xi r_0)J_k[\xi(R+r_0)]\xi d\xi \nonumber \\
&+&\int_{\xi=0}^\infty\sum_{k=-\infty}^{+\infty}J_k(\xi r_0)J_k[\xi(R+r_0)]f(\xi)\xi d\xi
\end{eqnarray}
Here $a(\xi)$ is again the redefined form.
As before, from the arbitrariness of $\xi d\xi$, we have
\begin{equation}
\sum_{k=-\infty}^{+\infty}J_k[\xi(R+r_0)]\left [a(\xi)\left
(1+\frac{\lambda^2}{\xi^2}\right)^{1/2}-J_k(2\xi r_0)-f(\xi)J_k(\xi r_0)\right]=0
\end{equation}
Hence we can write
\begin{equation}
a(\xi)=F(R,r_0,\xi)+f(\xi)
\end{equation}
where
\begin{equation}
F(R,r_0,\xi)=\frac{\sum_{k=-\infty}^{+\infty} J_k^\prime[\xi(R+r_0)]J_k(2\xi
r_0)}{\sum_{k=-\infty}^{+\infty} J_k^\prime[\xi(R+r_0)]J_k(\xi r_0)}
\end{equation}
and similarly
\begin{equation}
a(\xi)\left (1+\frac{\lambda^2}{\xi^2}\right)^{1/2}=G(R,r_0,\xi)+f(\xi)
\end{equation}
where
\begin{equation}
G(R,r_0,\xi)=\frac{\sum_{k=-\infty}^{+\infty}J_k[\xi(R+r_0)]J_k(2\xi r_0)}{\sum_{k=-\infty}^{+\infty}J_k[\xi(R+r_0)]J_k(\xi r_0)}
\end{equation}
Which further gives 
\begin{equation}
\frac{F+f}{G+f}=\frac{1}{\left (1+\frac{\lambda^2}{\xi^2}\right)^{1/2}}
\end{equation}
solving for $f(\xi)$, we have
\begin{equation}
f(\xi)=\frac{F\left (1+\frac{\lambda^2}{\xi^2}\right)^{1/2}-G}{1-\left
(1+\frac{\lambda^2}{\xi^2}\right)^{1/2}}
\end{equation}
Now using the identity
\begin{equation}
J_k^\prime(u)=\frac{1}{2}[J_{k-1}(u)-J_{k+1}(u)]~~{\rm{and}}~~J_\nu(v-u)=\sum_{-\infty}^{+\infty}
J_{\nu+k}(v)J_k(u)
\end{equation}
we have 
\begin{equation}
F=\frac{J_1[(R-r_0)\xi]}{J_1(R\xi)}~~{\rm{and}}~~ G=\frac{J_0[(R-r_0)\xi]}{J_0(R\xi)}
\end{equation}
Then substituting $f(\xi)$ in the expression for $\phi^{(s)}(r,z)$ (eqn.(27)), we have 
\begin{equation}
\phi^{(s)}(r,z)=e\int_{\xi=0}^\infty J_0(\xi
r)\left [\frac{F(1+\frac{\lambda^2}{\xi^2})^{1/2}-G}{1-\left
(1+\frac{\lambda^2}{\xi^2}\right)^{1/2}}\right] \exp(-\xi z) d\xi
\end{equation}
The corresponding electro-static field at $(R,z=0)$ is given by
\begin{equation}
E^{(s)}(R,0)=-\frac{\partial\phi^{(s)}(r,z)}{\partial r}=e\int_{\xi=0}^\infty \frac{J_1[(R-r_0)\xi]\left
(1+\frac{\lambda^2}{\xi^2}\right)^{1/2}-
\frac{J_0[(R-r_0)\xi]J_1(R\xi)}{J_0(R\xi)}}{1-\left (1+\frac{\lambda^2}{\xi^2}\right)^{1/2}} \xi d\xi
\end{equation}
where we have substituted $F$ and $G$ from eqn.(40).
The work function associated with the emission of electrons in the transverse direction 
is then given by \cite{R1}
\begin{equation}
W_f=-e\int_R^\infty E^{(s)}(R,0)dr_0
\end{equation}
Using the relation 
\begin{equation}
\int_0^\infty J_n(u)du=1
\end{equation}
we have
\begin{equation}
\int_R^\infty J_0[(R-r_0)\xi]dr_0=\int_R^\infty J_1[(R-r_0)\xi]dr_0=\frac{1}{\xi}
\end{equation}
Then finally we get 
\begin{equation}
W_f=-e^2\int_{\xi=0}^\infty \frac{\left (1+\frac{\lambda^2}{\xi^2}\right)^{1/2}-\left  [\frac{J_1(R\xi)}
{J_0(R\xi)}\right]} {1-\left (1+\frac{\lambda^2}{\xi^2}\right)^{1/2}}d\xi
\end{equation}
To evaluate $W_f$, let us consider part of this integral, given by 
\begin{equation}
I=-\int_0^\infty \frac{\left (1+\frac{\lambda^2}{\xi^2}\right)^{1/2}}{1-\left
(1+\frac{\lambda^2}{\xi^2}\right)^{1/2}} d\xi
\end{equation}
which is the first part of work function integral (eqn.(46)).
To evaluate integral $I$, let us put $\xi=\lambda \sinh \theta$ and then it is trivial to show that 
\begin{equation}
I=\frac{\lambda}{4}\int_0^\infty \frac{[\exp(\theta)+\exp(-\theta)]^2}{\exp(-\theta)} d\theta
\end{equation}
Obviously, the numerical value of this integral is infinity.

The other part of the work function integral (eqn.(46)) is also diverging in nature, but these
two divergences will not cancel each other. It is also quite obvious that in the limiting
cases, $R\longrightarrow \infty$, or $R\longrightarrow 0$, 
the diverging nature of work function for both the cases indicate that it is independent of the 
transverse dimension of the cylindrical cells. It is obvious that for 
$R\longrightarrow 0$, the second term on the numerator of eqn.(46) vanishes, 
whereas the rest can be integrated analytically and found to be diverging in the upper limit. 
On the other hand, for $R\longrightarrow
\infty$, the second term becomes $\sim \tan(\xi R-\pi/4)$ (asymptotically). Which itself also diverges
for certain values of the argument, but again dose not cancel with the infinity from the other part of this
integral. Therefore in this limit also the work function becomes
infinitely large. Such diverging nature of work function is associated with electron emission along
the transverse direction to the external magnetic field. Although we have considered
here the extreme case of ultra-strong magnetic field, we do expect that even for low and moderate 
field values, the work function corresponding to the electron emission in the transverse direction 
will be several orders of magnitudes larger than the corresponding longitudinal values (see also
\cite{pl,rs}).
\section{Conclusion}
In conclusion, we mention that the main purpose of this article is to show that (i) in presence of
strong quantizing magnetic field the work function becomes anisotropic, (ii) the transverse part is
infinitely large, (iii) the longitudinal part is finite but increases with the strength of magnetic field and finally, (iv) low
field values of work functions are more or less consistent with the tabulated values.
In the present work, with such simple minded model, to the best of our knowledge, 
the anisotropic nature of work function in presence of strong quantizing magnetic field 
is predicted for the first time. Further, the diverging character of work function associated 
with the electron emission in the transverse direction is also obtained for the first time, 
and so far our knowledge is concerned, it has not been reported in any published work.
We have also noticed that in the low magnetic field limit (within the limitation of this
model) the numerical values of work function are of the same orders of magnitude with the known
laboratory data.
However, we are not able to compare our results with the variation from one metal to another. 
The anisotropic nature of work function is apparently coming from the deformed cylindrical 
nature of electron distribution around iron nuclei caused by strong magnetic field at the outer crust region of a strongly
magnetized neutron star. 
Now because of Ohmic decay, when the field strength will become 
low enough $(<B_c^{(e)})$, we do expect that the spherically symmetric nature of the electron distribution around the nuclei will be restored. 
As a consequence, the electron emission processes will no longer be affected by strong magnetic field 
through work function of the metallic outer crust. If the diverging nature of work function
associated with the emission of electrons through the curved faces of the 
so called metallic atoms (they are in reality the electron distribution around 
the iron nuclei),
is solely because of cylindrical deformation, which in the present model caused by the presence of 
ultra-strong magnetic field, we therefore expect that for the emission of electrons through 
the curved surface of any object having cylindrical structure, the transverse component of
work function will be much larger compared to the longitudinal part. As a consequence the
electron emission current along the axial direction will be large enough compared to the
transverse emission current \cite{pl}. The work function associated with the emission of
electrons along the direction of magnetic field may be used to obtain electron field
emission current at the polar region of strongly magnetized neutron stars, which will help
us to understand the formation of magneto-spheric plasma. 

\section{Acknowledgement} We are thankful to Professor D. Schieber and Professor L. Schachter of Haifa 
for providing us an online re-print of \cite{R1}. 

\begin{figure}[t]
\vspace*{2mm}
\begin{center}
\includegraphics[width=8.3cm]{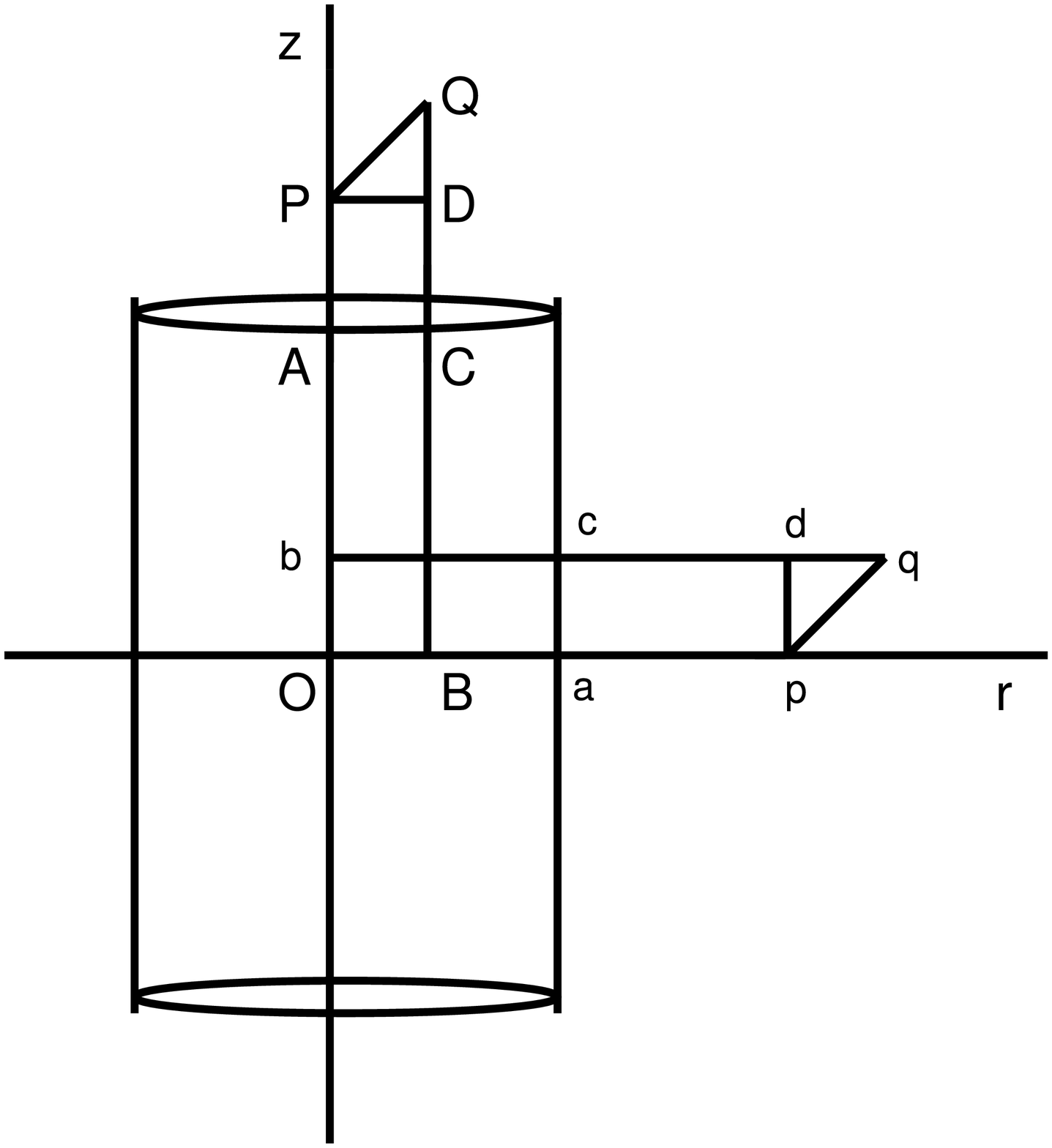}
\end{center}
\caption{
Schamatic diagram for electron emission along and orthogonal to the 
direction of magnetic field.
}
\end{figure}
\begin{figure}[t]
\vspace*{2mm}
\begin{center}
\includegraphics[width=8.3cm]{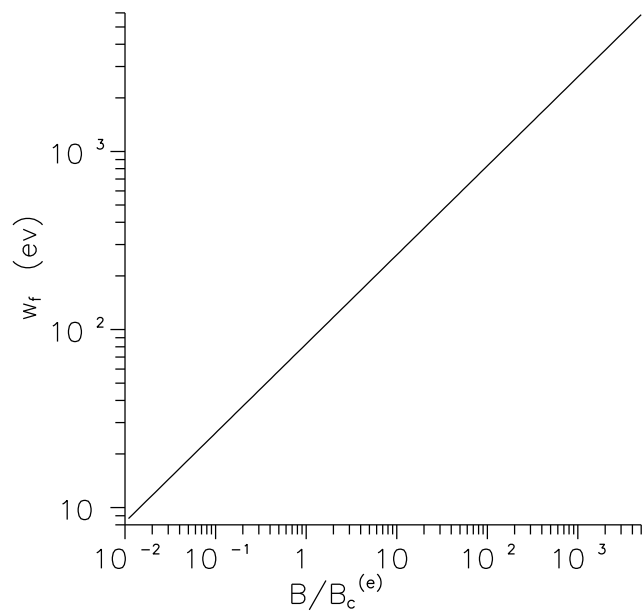}
\end{center}
\caption{
Variation of work function ($W_f$) with the magnetic field strength.
}
\end{figure}

\begin{thebibliography}{99}
\bibitem{R1} D. Schieber, Archiv f{$\ddot{\rm{u}}$}r Elektrotchnik, 67, 387, (1984).
\bibitem{sn} A. Sommerfeld, Naturwiss., 15, 825, (1927); 16, 374, (1928).
\bibitem{fr} J. Frenkel, Zeits. f. Physik 49, 31, (1928).
\bibitem{tb} J. Tamm and D. Blochinzev, Zeits. f. Physik 77, 774, (1932).
\bibitem{sk} J. C. Slater and H. M. Krutter, Phys. Rev. 47, 559, (1935).
\bibitem{ej1} E. Wigner and J. Bardeen, Phys. Rev. 48, 84, (1935).
\bibitem{jb} J. Bardeen, Phys. Rev. 49, 653, (1936).
\bibitem{R2} A. Jessner, H. Lesch and T. Kunzl, APJ, 547, 959, (2001). 
\bibitem{R3} M.A. Ruderman and P.G. Sutherland, APJ, 196, 51, (1975).
\bibitem{R4} see also, D.A. Diver, A.A. da Costa, E.W. Laing, 
C.R. Stark and L.F.A. Teodoro, astro-ph/0909.3581. 
\bibitem{R5} S.L. Shapiro and S.A. Teukolsky, Black Holes, White Dwarfs
and Neutron Stars, John Wiley and Sons, New York, (1983).
\bibitem{R6} F.C. Michel, Rev. Mod. Phys., 54, 1, (1982); F.C. Michel, 
Advances in Space Research, 33, 542, (2004).
\bibitem{R7} A.K. Harding and D. Lai, Rep. Prog. Phys., 69, 2631, (2006).
\bibitem{R8} M. Ruderman, Phys. Rev. Letts., 27, 1306, (1971).  
\bibitem{R12} R.H. Fowler and L.W. Nordheim, Proc. Roy. Soc. London A, 
119, 173, (1928).
\bibitem{ek} W. Ekardt, Phys. Rev. 29, 1558, (1984).
\bibitem{wc} L. L. Wang and H. P. Cheng, Euro. Phys. Jour. D, 43, 247, (2007)
\bibitem{mvvv} M. V. Mamonova and V. V. Prudnikov, Russian Physics Journal, 
41, 1174,(1998).
\bibitem{hph} Hennie C. Mastwijk, Paul V. Bartels, and Huub L. M. Lelieveld,
arXiv:0704.3797.
\bibitem{pl} for anisotropic nature of electron emissions from the tips and 
the walls of a Carbon nano-tubes, see P. Liu. et. al., Nano Letters, 8, 647, 
(2008).
\bibitem{rs} for the anisotropic nature of electron emissions from different 
crystal planes, see R. Smoluchowski, Phys. Rev. Letts., 60, 661, (1941).
\bibitem{R9} E.H. Lieb, J.P. Solovej and J. Yngvason, Phys. Rev. Letts., 69, 
749, (1992).
\bibitem{R10} V. Canuto and J. Ventura, Fundamentals of Cosmic Physics, 2, 203, (1977).
\bibitem{R11} Nandini Nag, Sutapa Ghosh and Somenath Chakrabarty, Ann. Phys.,
324, 499 (2009); Nandini Nag and Somenath Chakrabarty, Euro. Phys. Jour. A
{\bf{A45}}, 99, (2010).
\bibitem{pbh} A.Y.Potekhin, D.A.Baiko, P.Haensel and D.G.Yakovlev, Astron. Astrophys. 346, 345, (1999).
\bibitem{po} A.Y.Potekhin, Astron. Astrophys. 306, 999, (1996).
\bibitem{py} A.Y.Potekhin and D.G. Yakovlev, Astron. Astrophys. 314, 341, (1997).
\bibitem{md1} Z.Medin and D.Lai, MNRAS, 382, 1833, (2007).
\bibitem{md2} Z.Medin and D.Lai, Phys. Rev. A74, 062507, (2006).
\bibitem{ad} A.Harding and D.Lai, astro-ph/0606674.
\bibitem{lk} for the original work on work function using DFT, see N.D.
Lang and W. Kohn, Phys. Rev. B1, 4555, (1970); N.D. Lang and W. Kohn,
Phys. Rev. B3, 1215, (1971).
\end{thebibliography}
\end{document}